\documentclass[aps,prl,reprint,groupedaddress,showpacs,showkeys]{revtex4-1}

\usepackage{graphicx}
\usepackage{dcolumn}
\usepackage{bm}
\usepackage{amsmath}         
\usepackage{lineno}
\usepackage{xcolor}

\usepackage{hyperref}

\begin{document}

\preprint{}

\title{Elastic Properties of Symmetric Liquid-Liquid Interfaces }

\author{Ramanathan Varadharajan}
\email[R. Varadharajan: ]{ramanathan.varadharajan@wur.nl}
\author{Frans A. M. Leermakers}
\email[F. A. M. Leermakers: ]{frans.leermakers@wur.nl}
\affiliation{Physical Chemistry and Soft Matter, Wageningen University \& Research Center, Stippeneng 4, 6708 WE Wageningen, The Netherlands.}


\date{\today}

\begin{abstract}
 The mean ($\kappa$) and Gaussian ($\bar{\kappa}$) bending rigidities of liquid-liquid interfaces, of importance for shape fluctuations and topology of interfaces, respectively, are not yet established: even their signs are debated. Using the Scheutjens Fleer variant of the self-consistent field theory, we implemented a model for a symmetric L/L interface and obtained high precision (mean field) results in the grand canonical $(\mu, V, T)$-ensemble. We report positive values for both moduli when the system is close to critical where the rigidities show the same scaling behavior as the interfacial tension $\gamma$. At strong segregation, when the interfacial width becomes of the order of the segment size, $\bar{\kappa}$ turns negative. The length scale $\lambda \equiv \sqrt{\kappa/\gamma}$ is of order the segment size for all strengths of interaction; yet the $1/\sqrt{N}$ chain length correction reduces $\lambda$ significantly when the chain length $N$ is small.

\end{abstract}


\maketitle

Out-of-plane fluctuations of liquid-liquid (L/L) interfaces are controlled by the interfacial tension ($\gamma$) for wavelengths larger than the cross-over length $\lambda = \sqrt{\kappa/\gamma}$ and by the bending rigidity ($\kappa$) at shorter wavelengths. This $\lambda$ should be comparable to a molecular length scale \cite{laradji2000elastic}. Precise prediction of $\lambda$ from a molecular model would significantly advance our understanding on fluctuations in L/L interfaces. However, molecular models that have access to $\kappa$ at sufficient accuracy have not yet been forwarded. More specifically, a key issue here is that molecular theories thus far have failed to establish the sign of $\kappa$.

As $\kappa$ controls the magnitude of the fluctuations (at short length scales), we expect it to be positive ($\kappa>0$). In stark contrast to this, surprisingly few theoretical predictions foresee a positive value \cite{blokhuis1992derivation}: to date, molecular models typically predict negative values \cite{matsen1999elastic,leermakers2013direct,oversteegen2000rigidity,blokhuis1999fluctuation}. Nevertheless in (mesoscale) simulations \cite{van2002determination, muller2002elastic, muller2005incorporating} and in phenomenological models \cite{laradji1998elastic} a positive sign for $\kappa$ is often chosen. 

Besides $\kappa$, interfaces have a second elastic constant known as the saddle spay modulus or Gaussian bending rigidity $\bar{\kappa}$. The $\bar{\kappa}$ should control the topology of interfaces; a negative value will prevent the formation of saddle shaped interfaces whereas a positive value will promote these. A sign-change (e.g. upon a change of temperature) is easily envisioned. However, existing predictions indicate a strictly positive value \cite{matsen1999elastic,leermakers2013direct,oversteegen2000rigidity}.

Molecular theories give relatively easy access to the accurate values for the interfacial tension \cite{vanderWaalsThesis,van1979thermodynamic}. However, the evaluation of the rigidities has many intricacies. With respect to common practise, we found that a sound estimation of $\kappa$ and $\bar{\kappa}$ from a molecular theory require to overcome two hurdles: (i) to quantify the curvature expenses at a fixed chemical potential ($\mu$) of all molecular species and (ii) to properly account for non-local contributions to the enthalpic interactions. 

In this letter, we successfully overcome these theoretical challenges and show that $\kappa$ is strictly positive for L/L interfaces and hence fluctuations from planar state cost free-energy. We observe that $\lambda$ is of the order of the segment size in the limit of strong or weak segregation, yet shows a non-monotonous behavior in transition regime. We discuss the implications of chain length ($N$ - degree of polymerization) on fluctuations of L/L interfaces, in light of results obtained for $\lambda$. Finally, we present and discuss the sign-switch for $\bar{\kappa}$. 

Mean field results for a simple symmetric interface between two liquids $A_N$ and $B_N$, where $N$ is chain length (degree of polymerization), is discussed. The case with $N=1$ correspond to the well-known van der Waals interface \cite{van1979thermodynamic,safran1994statistical}. When $N$ is large we arrive at another well studied interface, namely between two inmiscible polymers \cite{semenov1993theory,matsen1999elastic}. As $\gamma>0$, the system has a tendency to minimize its area. Thermal energy causes the macroscopic interface to fluctuate. The accompanied entropy gain is counteracted by an unfavourable increase in area and a penalty for the interface to (locally) bend away from the planar ground state. Such curved interfaces cannot maintain their tension exactly. Following Helfrich \cite{helfrich1973elastic} we consider a Taylor series expansion of the tension in terms of the  mean ($J=1/R_1 + 1/R_2$) and Gaussian ($K=1/R_1 \times 1/R_2$) curvature ($R_1$ and $R_2$ are two principle radii of curvature): 
\begin{equation}
    \gamma(J,K)=\gamma(0,0) +\frac{\partial \gamma}{\partial J}J + \frac{1}{2}\frac{\partial^2 \gamma}{\partial J^2}J^2 + \frac{\partial \gamma}{\partial K}K + \cdots 
    \label{eq:helfrich}
\end{equation}
The term linear in $J$ is well documented and properly understood \cite{rowlinson2013molecular,mitrinovic2000noncapillary,mitrinovic2000noncapillary, wu1993surface, ocko1994x, merkl1997influence}. Here and below, we will focus on symmetric interfaces for which this linear term vanishes. Defining the mean bending modulus, $\kappa \equiv \frac{\partial^2 \gamma}{\partial J^2}$, and Gaussian bending rigidity, $\bar{\kappa} \equiv \frac{\partial \gamma}{\partial K}$, Eqn. \ref{eq:helfrich} reduces for weakly curved interface to $\gamma(J,K)-\gamma(0,0)=\frac{1}{2}\kappa J^2 + \bar{\kappa}K$. The grand potential, $\Omega = \int \gamma {\rm d}A$, quantifies the excess free energy of the interface and is the characteristic function that can be used to describe the fluctuations of the interface that take place at specified chemical potentials, $\mu$, (that of the binodal) of the molecular species in the system.

We implement a self-consistent field model using lattice approximations as introduced by Scheutjens and Fleer for polymer adsorption \cite{scheutjens1979statistical,scheutjens1980statistical}. These authors combined a freely jointed chain model with a Flory-Huggins equation of state. The repulsive interactions between $A$ and $B$ segments is quantified by a Flory-Huggins interaction parameter. When $\chi > \chi^c = 2/N$ the system features a solubility gap. It turns out that it is important to understand how the SF-SCF formalism deviates from the classical SCF theory that is used to describe microphase segregation (which is also frequently used to model the interface between two polymeric solutions). 

In SF-SCF we write a mean field free energy (in dimensionless units) for a molecularly inhomogeneous system \cite{leermakers2013bending,kik2010molecular,fleer1993polymers,evers1990statistical} in terms of volume fraction $\varphi_x({\bf r})$ and complementary segment potential $u_x({\bf r})$ profiles for segment types $x= A,\ B$ on a grid of lattice sites with characteristic size $b$ equal to segment size. To facilitate proper extremization we add a Lagrange parameters, $\alpha({\bf r})$, in free energy to implement the local incompressibility constraint, $\varphi_A(r)+\varphi_B(r)=1$, applicable in each coordinate $r$ and a parameter $\Delta_{r;r_0} \equiv \delta_{r;r_0}\nu$, where $\delta_{r;r_0}$ is unity when $r=r_0$ and zero otherwise and $\nu$ is the Lagrange parameter coupled to the requirement that at the interface location $r=r_0$ the density of both components match \cite{matsen1999elastic}:
\begin{equation}
\begin{aligned}
F = - \ln Q([u]) - \sum_{x,r}u_x(r)\varphi_x (r) L(r) + F^{\rm int}([\varphi]) +\\
\sum_{r}\alpha(r)\bigg{[}\sum_x\varphi_x(r) - 1\bigg{]}
+ \Delta_{r;r_0}\bigg{[}\varphi_A(r) -\varphi_B(r)\bigg{]}  
\label{eq.scff}
\end{aligned}
\end{equation}
In the mean field approach one can decompose the partition function $Q=\Pi_i q_i^{n_i}/n_i!$ where $i=1,\ 2$ refers to $A_N$ and $B_N$ respectively. The molecular partition function $q_i$ contains the statistical weight of all possible and allowed conformations of molecule $i$ (see below). $n_i$ is the number of molecules of type $i$ in the system. $L(r)$ gives the number of lattice sites at the lattice coordinate $r$. For planar system $L(r)=1$ is a constant (all quantities are per unit area), in cylindrical coordinates $L(r) \propto r$ (and quantities are expressed per unit length of the cylinder), while in spherical coordinates $L(r) \propto r^2$. The interaction free energy is a function of the densities:
\begin{equation}
F^{\rm int}=\chi\sum_r L(r) \varphi_A(r) \bigg{[} \langle \varphi_B(r)\rangle -\varphi_B^b \bigg{]}
\end{equation} 
Here $\varphi_B^b$ refers to the bulk volume fraction of B (of one of the bulk phases). Importantly, the angular brackets are needed to account for the contact energy in a system with gradients in density. Similar as in the Cahn Hilliard theory \cite{safran1994statistical} we write 
\begin{equation}
   \langle \varphi_B(r)\rangle= \varphi_B(r)+\frac{1}{6} \nabla^2 \varphi_B(r)
   \label{eq:site}
\end{equation}
where the $\nabla^2$ is easily mapped on the lattice as explained extensively in earlier literature \cite{scheutjens1979statistical,leermakers2013bending}. SCF solutions now involve optimizing the free energy ($F$) with respect to its variables, respectively segment potentials, volume fractions. When $\partial F/\partial \varphi_A({r}) = 0$, we find that the potentials must obey $u_A({r})=\alpha(r)+\Delta_{r;r_0}+\chi (\langle \varphi_B(r) \rangle-\varphi_B^b)$ (and similarly for $u_B(r)$). Setting $\partial F/\partial u_x(r) =0$ shows the way to evaluate the densities: $\varphi_x(r)=-\partial \ln Q/\partial u_x(r)$. The molecular partition functions are found from the endpoint distribution $q_i=\sum_r G_i(r,N|1)$. The end-point distributions are recursively found from $G_i(r)=\exp -u_i(r)$ by the propagator $G_i(r,s|1)=G_i(r) \langle G_i(r,s-1|1)\rangle$, where the angular brackets have the same meaning as in Eqn.\ref{eq:site}. The segment densities are found by the composition law, which for homopolymers read: $\varphi_i(r) = C_i G_i(r,s|1)G_i(r,N-s-1|1)/G_i(r)$. As the position of the interface is already controlled by a Lagrange parameter $\Delta_{r;r_0}$, we no longer need to specify a fixed amount of one of the components (as is needed in a canonical ensemble), but we can normalize the densities with $C_i=\varphi_i^b/N_i$ where $\varphi_i^b$ is specified by the binodal: A proper binodal value is a (relevant) root of the Flory-Huggins Eqn. $\ln \frac{\varphi}{1-\varphi}+\chi N (1-2\varphi )=0$.

Numerical solutions, which routinely were obtained with an accuracy of 9 significant digits, that optimize the free energy functional and obey to all constraints, have the property that the potentials both determine and follow from the volume fractions profiles and \textit{vice versa} and are said to be self-consistent.  
For such solution one can compute the grand potential by $\Omega=\sum_r L(r) \omega(r)$ wherein the grand potential density at coordinate $r$ is given by 
$\omega(r) = -\sum_i (\varphi_i( {r})-\varphi_i^b)/N_i - \alpha( {r}) -
\chi(\varphi_A( {r})\big{[}\varphi_B( {r}) + \frac{1}{6} \nabla^2 \varphi_B( {r})\big{]}-\varphi_A^b \varphi_B^b)$. The planar interface has a tension $\gamma_p = \sum_z \omega(z)$, where $z$ is the coordinate in the planar system. This planar interface serves as the ground state or reference state needed to estimate the grand potential increase of the curved interfaces.

SCF solutions are routinely created for planer (p) cylindrical (c) and spherical (s) coordinates. As the position of the interface is exactly known and specified by $\Delta_{r;r_0}$ to be at coordinate $r = r_0$ we obtain the interfacial tensions in all cases unambiguously. In spherical geometry we have $R\equiv R_1=R_2=r_0-1/2$, while in cylindrical geometry $R\equiv R_1=r_0-1/2$ and $R_2=\infty$ and  $\gamma_s=\Omega_s/(4 \pi R^2)$ and $\gamma_c=\Omega_c/(2\pi R)$ for spherical and cylindrical geometries, respectively. Here $\Omega_s$ is the grand potential when the interface is curved in a spherical fashion, and $\Omega_c$ is the grand potential per unit length of the interface when curved in a cylindrical fashion. Here we have implemented that a lattice site at coordinate $r$ is a distance $r-1/2$ away from the center of the coordinate system. In all calculations we make sure that the numerical value for $r_0$ significantly exceeds the width of the interface. Next we compute $\gamma_c-\gamma_p$ as well as $\gamma_s-\gamma_p$  that use the Helfrich equation \ref{eq:helfrich} to extract with high accuracy both $\kappa$ (from cylindrical geometry) and $2\kappa +\bar{\kappa}$ (from spherical geometry). The combination  of these results leads to both $\kappa$ and $\bar{\kappa}$. Note that in all calculations, $\mu$ for the molecular species were set to the value at the appropriate binodal. Invariably, we find a positive value for the mean bending modulus whereas the sign of $\bar{\kappa}$ is (as expected) not fixed. 

Results for $N=1$ (the van der Waals interface\cite{vanderWaalsThesis,van1979thermodynamic}) are presented in SI, here we focus on the captivating results for $N>1$ and understand that $N=1$ is a limiting case. 

\begin{figure}
\centering
\includegraphics[width=0.45\textwidth]{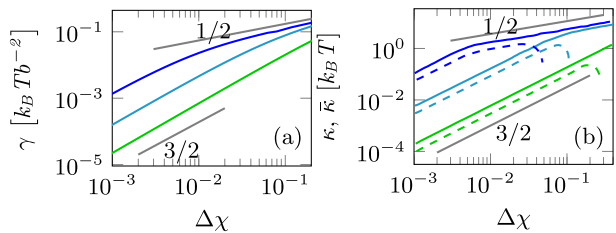}
\caption{(a) Interfacial tension per unit area $\gamma$ in units of [$k_BTb^{-2}$] as a function of $\Delta\chi=\chi  - \chi^{\rm c}$ in log-log coordinates (b) Mean $\kappa$ (solid line) and Gaussian bending rigidities $\bar{\kappa}$ (dashed line) in units of [$k_BT$] as a function of $\Delta\chi$. Relevant slopes are indicated. Three different chain lengths were used ($N=2:$ green, $N=20:$ cyan, $N=200:$ blue). }
\label{f3}
\end{figure}
In Fig. \hyperref[f3]{1(a)}, we present the interfacial tension and in Fig. \href{f3}{1(b)} the bending rigidities, both as a function of $\Delta \chi$, where $\Delta \chi \equiv \chi-\chi^c$ \textcolor{black}{($\chi^c = 2/N$: bulk critical point)}, for three values of the chain lengths $N = 2,\ 20$ and $200$. The corresponding results for the density difference and the interfacial width are presented \cite{supplemental}. The results for the interfacial tension are in principle well known \cite{safran1994statistical,helfand1971theory,helfand1975theory,matsen1999elastic,semenov1993theory,semenov1994scattering}. As long as the interface is wide compared to the coil size we find $\gamma \propto N (\Delta \chi)^{3/2}$ and in the other limit, where the interfacial width is small compared to the coil size, we have $\gamma \propto (\Delta \chi)^{1/2}$. (The 3/2 exponent is the mean field prediction, known to be subject to changes, the 1/2 exponent is expected to be accurate.) As in this regime $\chi$ exceeds \textcolor{black}{$\chi^{c}$} a lot, the result is similar to the known result that $\gamma \propto \sqrt{\chi}$. The latter result/regime is referred to as the \textit{strong segregation} limit and the former regime will be referred to as the \textit{weak segregation} limit. 

Interestingly the results for both bending rigidities [cf. Fig. \href{f3}{1(b)}] follow the results for the interfacial tension qualitatively. In the weak segregation the $3/2$ scaling is found, while for the strong segregation the $1/2$ scaling is recovered. It is important to mention that the Gaussian bending modulus deviates from the latter power-law dependence rather abruptly: quite suddenly the Gaussian bending rigidity goes to zero and then changes its sign. We will discuss this behaviour below in more detail. Comparing Figs. \href{f3}{1(a)} and \href{f3}{1(b)} shows that in the transition regions between weak and strong segregation the tension and rigidities do not exactly copy their dependencies. This has interesting implications as we will show next. 
\begin{figure}
    \centering
\includegraphics[width=0.4\textwidth]{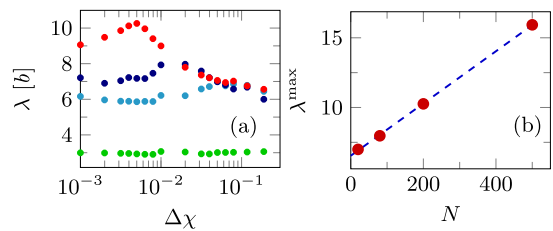}
\caption{$\lambda$ as a function of $\Delta\chi$ for different $N = 2$ (green), 20 (light blue), 80 (dark blue), 200 (red). \textcolor{black}{Guide lines for 0.1$\%$ uncertainty in $\kappa$ are provided to visualize the trend, points show original data}. 
(b) $\lambda^{\rm max}$ vs $N$ showing a linear dependence of the maximum cross-over length with respect to the chain length for ($N>20$). Dots indicate extracted values, blue dashed line is a linear fit $\lambda^{\rm max} = 0.02N + 6.54$ .  }
\label{f4}
\end{figure}

Above we introduced the length scale $\lambda=\sqrt{\kappa/\gamma}$. Combining results form Fig. \href{f3}{1(a), 1(b)}, we present $\lambda$ as a function of $\Delta \chi$ in semi-logarithmic coordinates for a few systems that differ in $N$ in Fig. \href{f4}{2(a)}. In this figure it can be seen that $\lambda$ goes through a local maximum \textcolor{black}{($\lambda^{\rm max}$)} in-between weak and strong segregation. Further, $\lambda$ reaches some fixed value, which is approximately 16\% larger at weak-, compared to strong segregation. In Fig. \href{f4}{2(b)}, linear dependence of $\lambda^{\rm max}$ is presented as a function of chain length $N$. Computer simulations that are aimed to find the bending rigidity from the height fluctuations of the interfaces \cite{muller1995structural,van2002determination} may benefit from relatively large $\lambda$-values and preferably should be executed for large chains, because $\lambda^{\rm max}$ grows linearly with $N$ as $\lambda^{\rm max} = 0.02N + 6.54$ . 

It is of interest to consider the ratio $\bar{\kappa}/\kappa$. Obviously, when $\bar{\kappa}$ is switching its sign, we cannot have a fixed ratio between the rigidities, but slightly outside the transition regions, that is both at weak and at strong segregation, the ratio is remarkably constant. This is illustrated in Fig. \href{f5}{3(a)} where for a selected number of values for $\Delta \chi$-values this ratio is plotted as a function of the chain length $N$. For low values of $N$ and low values of $\Delta \chi$ we have the typical weak segregation result. It could have been concluded from Fig. \href{f3}{1(b)}, the ratio goes to a constant of approximately $1/2$. In the high $N$ high $\Delta \chi$ value limit, that is, the strong segregation limit, where the Gaussian bending rigidity is negative, the ratio is approximately $-3/2$.  Implicit in this prediction is that also in the regime where $\bar{\kappa}<0$, the (negative) Gaussian bending rigidity follows a scaling law: $-\bar{\kappa}\propto (\Delta \chi)^{1/2}$.

Recently, we have shown for microemulsion systems that the Gaussian bending rigidity is positive for systems near the (bulk) critical point and negative otherwise \textcolor{black}{\cite{varadharajan2018sign}}.  From Figs. \href{f3}{1(b)} and \href{f5}{3(a)}  it can be seen that for liquid/liquid interfaces the same phenomenology applies and it is of interest to quantify this sign switch. It is mentioned in SI \cite{supplemental} that the scaling for the interfacial width $W\propto (\Delta \chi)^{-1/2}$ applies both at weak and strong segregation regimes. The appropriate prefactor depends on $1+1/N$ in weak segregation and on $1+1/\sqrt{N}$ in strong segregation regime \cite{supplemental,mocan2018microphase}. In Fig. \href{f5}{3(b)} we present a diagram of states in coordinates $\sqrt{\Delta \chi}$ vs $1/\sqrt{N}$. The dashed line ($(\Delta\chi)^{1/2} = 0.2+0.5/\sqrt{N}$) separates the parameter combination with positive $\bar{\kappa}$ values from the negative ones. The line in Fig. \href{f5}{3(b)} is functionally consistent with the prefactor for the interfacial width scaling in the strong segregation limit and therefore we speculate that the interfacial width controls the sign switch. Apparently, the Gaussian bending rigidity changes sign when the width of the interface is approximately 3 to 4 times the segment size. This thus happens at relatively weak segregation for short chains but in the strong segregation regime for long chains.   
\begin{figure}
\centering
\includegraphics[width=0.45\textwidth]{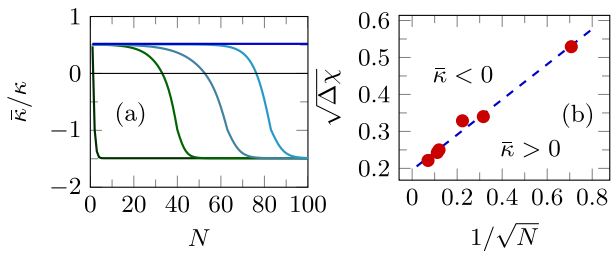}
 \caption{(a) Ratio of Mean and Gaussian bending rigidities as a function of molecular weight $N$ at $\Delta\chi = 0.03$ (dark blue), 0.06 (light blue), 0.075 (cyan), 0.08 (green), 0.3 (black) showing transition from $\bar{\kappa}/\kappa= 1/2 \to -3/2$ as the system is going from weak to strong segregation by increasing the chain length. (b) Diagram of states for the sign switch of $\bar{\kappa}$. The dashed fit-line is explained in the text.}
 \label{f5}
\end{figure}

Physically, the implication of a positive $\kappa$ is that fluctuations of the L/L interface away from the planar state do cost (free)energy. In the light of existing literature, this expected result is remarkable for several reasons. 

(i) The way the interface is pinned, using a Lagrange parameter coupled to the equal density of the two liquid component,  was first used by Matsen \cite{matsen1999elastic}. He implemented this method to find the bending rigidities using the classical self-consistent field machinery. His SCF approach has many similarities with the (current) SF-SCF approach. Yet he reported strictly negative values for $\kappa$ and positive values for $\bar{\kappa}$. The only relevant difference between our SF-SCF approach \textcolor{black}{and} the classical SCF approach of Matsen rests in the fact that for the interactions we have implemented the Cahn-Hilliard gradient terms, and Matsen did not. In the SI we show that when in SF-SCF these gradient terms are neglected, that is, when we implement (cf Eqn \ref{eq:site}) $\langle \varphi(r)\rangle \to \varphi(r)$, we do reproduce all results of Matsen to a high accuracy. This proves the importance of the non-local interaction contributions to determine the rigidities. In the absence of these non-local interactions, neither the sign nor the scaling dependencies are apparently properly predicted. 

(ii) A number of years ago Blokhuis has shown that a big effect on how bending of the interface is implemented \cite{blokhuis2009spectrum} must be expected. He identified the so-called equilibrium bending mode where $\mu$ controls the curvature. In this case, the bending of the interface is accompanied by the development of a Laplace pressure inside the `droplet' phase. $\gamma$ is then typically computed at the so-called surface of tension (SOT). The position of the interface is taken to be at the SOT, even though other choices can be implemented. When the tension evaluated at this SOT is used in the Helfrich equation one can evaluate $\kappa$ and $\bar{\kappa}$ (again using the combination of cylindrical and spherical geometries). As confirmed by SF-SCF calculations \cite{leermakers2004negative,oversteegen2000rigidity}, in this case $\kappa$ is negative and $\bar{\kappa}$ is strictly positive. Also for equilibrium bending one finds scaling behaviour for both moduli when the system is close to the bulk critical point. However in this case the coefficient of $1/2$ is found. Blokhuis also analyzed the bending of the interface at fixed $\mu$ (binodal values) by controlling the position of the interface by some (local) external field. Interestingly, in this case he recovered the 3/2 scaling law, similarly as presented above for the weak segregation \cite{blokhuis2008description}. However, still the value of $\kappa$ was negative. Interestingly, quantitative values for $\kappa$ did depend on the choice that was implemented to define the interface position. Blokhuis could not exclude that there might be some choices for this that could turn $\kappa$ positive. Hence the current results that shows positive values for $\kappa$ and a 3/2 scaling coefficient (in the weak segregation limit) indeed is the anticipated result. 

Perhaps the more interesting prediction is the sign switch of $\bar{\kappa}$. In surfactant systems such sign switch has been found earlier \cite{varadharajan2018sign} and is expected because it correlates with the rich phase behavior for these systems that include cubic phases and sponge phases. For the liquid/liquid interface  the sign switch of $\bar{\kappa}$ is unknown. It will be of more than average interest to find experiments for which this sign switch is important. This is not trivial because we know that the prime interest of an interface is to reduce its area under the influence of a finite tension. However, when the interfaces are strongly perturbed, one might find that drops may pinch off. Arguably this is easier when $\bar{\kappa} <0$, hence at strong segregation systems, and suppressed otherwise. This reasoning may explain why for some liquids one can manipulate the splashing by an external pressure \cite{xu2005drop,yarin2006drop}. Such effects may find applications in various industrial process \cite{tekin2008inkjet,duez2007making}. Our results may also have implications in emulsion droplet formation as the ease by which drops form might be manipulable by the sign and size of $\bar{\kappa}$. 

We have proved that the fluctuations from L/L interface away from the planar interface indeed cost free energy. We have shown that the cross-over length has a non-monotonous behavior in the transition regime between weak and strong segregation. Besides this, $\lambda$ is essentially constant (of the order of a few segments lengths) and does not vary much with chain length and/or distance to a critical point. Moreover, a sign-switch of $\bar{\kappa}$ is now established. As interfaces are omnipresent, it is difficult to overestimate the many implications of our phenomenal results which may include complex phenomena such as droplet nucleation from a supersaturated solution, emulsion formation and wetting phenomena to mention a few.

This work is part of an Industrial Partnership Programme, `Shell/NWO Computational Sciences for Energy Research (CSER-16)', of the Foundation for Fundamental Research on Matter (FOM), which is part of the Netherlands Organization for Scientific Research (NWO). Project number: 15CSER26.

\bibliography{main}
\end{document}


\preprint{}

\title{Supplementary information for Elastic Poperties of Symmetric Liquid-Liquid Interface}

\author{Ramanathan Varadharajan}
\email[R. Varadharajan: ]{ramanathan.varadharajan@wur.nl}
\author{Frans A. M. Leermakers}
\email[F. A. M. Leermakers: ]{frans.leermakers@wur.nl}
\affiliation{Physical Chemistry and Soft Matter, Wageningen University \& Research Center, Stippeneng 4, 6708 WE Wageningen, The Netherlands.}


\date{\today}



\pacs{}

\maketitle
The equilibrium properties of symmetric L/L interfaces is solved in a mean field model. Solutions of the SF-SCF equations lead not only to thermodynamic information on interfaces but also provides insight in corresponding structural properties. In the first section of this supplementary information we illustrate this by focusing on a few density profiles across symmetric interfaces and then illustrate how curvature of the interface changes the interfaces in a subtle yet expected ways. In a second paragraph we  present result for the thermodynamic and elastic properties of interfaces between monomeric solvents (the van der Waals case). In a third section the finite chain length corrections of the interfacial tension as well as the mean bending rigidity are analysed to underpin how the length scale $\lambda = \sqrt{\kappa/\gamma}$ has a non-monotonous behaviour in-between the weak and strong segregation regimes. The fourth section gives information on the interfacial width and the density difference across the interface as functions of the interaction parameter. The final section is devoted to the comparison of the SF-SCF results with the classical results of Matsen \cite{matsen1999elastic}. In this comparison we modified the SF-SCF approach by ignoring the non-local interactions in the segment potential. The match of results of these two approaches proves that the difference found between full SF-SCF and the classical SCF used in polymeric interfaces must be attributed to these non-local interactions and not to, e.g., spurious numerical issues.  

\subsection{On the segment density profiles across symmetric interfaces and how curvature of the interface modulates these.}
The SF-SCF theory has been outlined in the paper. A key result of the theory is the structure of the interface. This interface readily forms both for monomeric systems $A_1$-$B_1$ (referred to as the van der Waals interface), as well as for polymeric interfaces $A_N$-$B_N$ where $N$ is the degree of polymerisation. The reason why in monomeric systems the interface is stabilized is due to the non-local interactions in the segment potential. For polymer systems the interface in principle should also experience such Cahn-Hilliard gradient contributions (leading to a $k (\Delta \varphi)^2$ in the Landau free energy), but they are already stabilized by a Lifshitz-Edwards entropy ($b^2(\Delta \sqrt\varphi)^2/12$ in the Landau free energy \cite{fleer1993polymers}). Because of the latter effect, in polymeric models the Cahn-Hilliard contribution is often ignored. We show that this Ansatz is leading to wrong predictions for the bending rigidities and this is in more depth elaborated on in the final section of this supplementary information. 

\begin{figure}[htpb!]
    \centering
    \includegraphics[width=\textwidth]{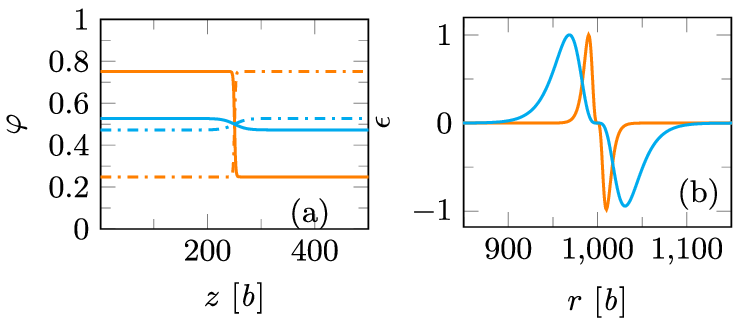}    
    \caption{(a) Volume fraction profile ($\varphi_A$: solid, $\varphi_B$: dashed)for planar liquid-liquid interface at strong segregation ($\Delta\chi=0.1$, Orange) and weak segregation ($\Delta\chi=0.001$, blue). (Note $\chi^{\rm c} = 2$) (b) Corresponding normalized grand potential density for spherically curved interface $\epsilon$ as a function of radial distance. $\epsilon(r) = \frac{\omega_s(r)-\omega_p(z)}{\max |\omega_s(r)-\omega_p(z)}$. Interfaces are pinned at $r=1000$ and $z=1000$, respectively.}
    \label{f1}
\end{figure}

In Fig. \ref{f1}a the volume fraction profiles are presented for simple planar monomeric Liquid-Liquid interface at strong segregation (orange) and weak segregation (cyan). Interfacial profiles converge to the classical result $\varphi(z)= 0.5 \pm \Delta \varphi \tanh((z-z^*)/W)$ in the limit of weak segregation. Here $z=z^*$ is the position of the interface, $W=b \sqrt{\lambda \chi/\Delta \chi }$ is the width of the interface ($\lambda =1/6$, $b$ is the size of a segment), and $\Delta \varphi$ is the density difference (between binodal concentrations) which near critical is given by $\sqrt{6 \Delta \chi /16}$. Fig. \ref{f1}a is instrumental in visualizing the length scales involved in the problem. Primary length scale, as observed from the figure is the interfacial width $W$. It is clear that closer to the bulk critical point the interfacial width widens and the density difference decreases as observed \cite{safran1994statistical}. Other length scales associate are the coil size $R_g=b\sqrt{N/6}$  and the segment size $b$. We note that these length scale will play a significant role in polymeric liquid-liquid interfaces and are addressed in the paper. 

The mechanical parameters of the interface are found by comparing the grand potential of the planar interface with those found in cylindrical or spherical geometry. We use the coordinate $r$ for non-planar geometries and note that $r=0$ is the center coordinate. Upon curving the interface small deviations for the $\tanh$-profiles occur. As a result also the grand potential density profile $\omega(r)$ deviates from the corresponding $\omega_p(z)$. In the planar interface $\omega_p(z)$ has a maximum positive value at the point where $\varphi(z)=0.5$, i.e. at $z=z^*$, and the grand potential density decays exponentially to zero with length scale $\xi$. This function is exactly symmetric meaning that $\omega(z^*-a)=\omega(z^*+a)$ for all values of $a$. In curved geometries this is no longer the case as illustrated in Fig. \ref{f1}b. The grand potential density difference $\epsilon(r)$ (see the legend for the definition of this difference; the absolute value of $\omega$ is a strong function of $\Delta \chi$ and therefore we normalised the grand potential density differences so that a comparison is possible) has a positive excursion for $r$-values smaller than the interface position $r=r^*$ (here $r^*=1000$) and a negative excursion for larger $r$.  In the weak segregating case the width of the interface is wide and therefore the  $\epsilon(r)$-profile varies further away from the position of the interface as well.  

\subsection{Interfacial tension and bending rigidities of monomeric L/L interface}

\begin{figure}[htpb!]
\centering
\includegraphics[width=\textwidth]{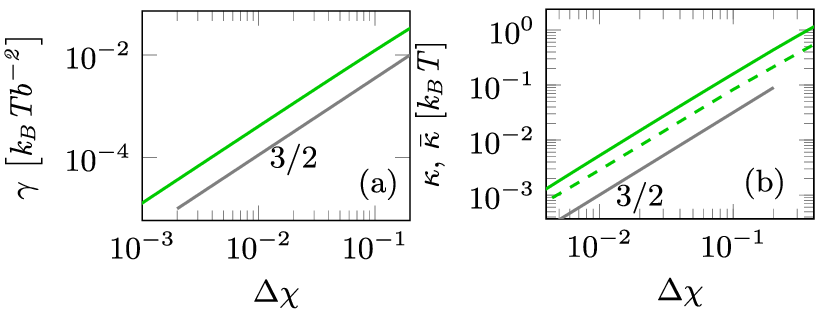}
\caption{(a) Interfacial tension $\gamma$ in units of [$k_BTb^{-2}$] as a function of $\Delta\chi=\chi  - \chi^{\rm c}$ in log-log coordinates (b) Mean $\kappa$ (solid line) and Gaussian bending rigidities $\bar{\kappa}$ (dashed line) in units of [$k_BT$] as a function of $\Delta\chi$. Relevant slopes are indicated. Results are presented for monomeric L/L interface. }
\label{sf2}
\end{figure}
In the article we have presented results for the interfacial tension $\gamma$ as well as the bending rigidities $\kappa$ and $\bar{\kappa}$ for symmetric interfaces with molecular weights $N=2$, $20$ and $200$ as a function of $\Delta \chi= \chi - \chi^{\rm c}$. Similar results for the $N=1$ system, that is for the van der Waals interface, are presented in fig. \ref{sf2}. As expected the results are extremely close to those presented in the text for $N=2$: In good approximation $\gamma \propto (\Delta \chi)^{3/2}$. The same power-law dependencies are found for the mean and Gaussian bending rigidities. For all values of $\Delta \chi$ that we can reliably evaluate in the current implementation of the SF-SCF equations (up-to $\chi \approx 2.3$), the Gaussian bending rigidity follows the mean bending rigidity. In other words, in this case we do not witness the sign switch for the monomeric system. However, we do expect that for for the van der Waals interface the Gaussian bending  will change its sign, but this should happen for $\chi>2.3$. The ratio between $\bar{\kappa}/\kappa \simeq 0.5$ for all values of $\Delta \chi$. A slight deviation in the power law scaling near $\Delta\chi = 10^{-3}$ is attributed to computation complications.

\subsection{Finite chain length effects for interfacial tension and the mean bending rigidity in the weak segregation regime}
\begin{figure}[htpb!]
\centering
\includegraphics[width=\textwidth]{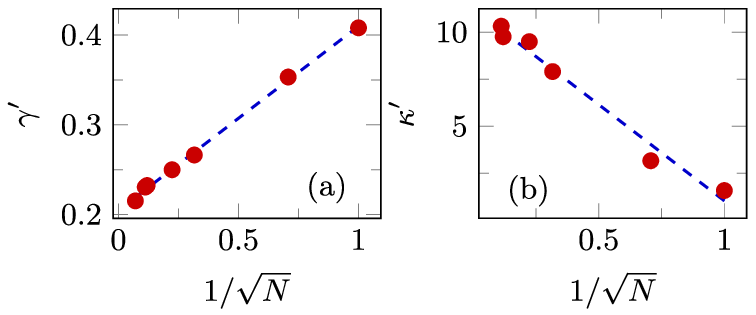}
\caption{(a) $\gamma'$ as a function of $1/\sqrt{N}$. Blue line is a linear fit 
$\gamma' \simeq (0.20 + 0.20/\sqrt{N})$
This implies at weak segregation regime $\gamma \simeq (0.20+0.20\frac{1}{\sqrt{N}}) N\Delta\chi^{3/2}$. (b) $\kappa'$ as a function of $1/\sqrt{N}$. Dashed blue line is a linear fit $\kappa' \simeq (11.26 - 10.22/\sqrt{N})$. This implies at weak segregation regime $\kappa \simeq (11.26-10.22\frac{1}{\sqrt{N}}) N\Delta\chi^{3/2}$ }
\label{sf3}
\end{figure}
In the main text it was shown that in weak segregation, that is for the systems where the interfacial width exceeds the coil size $R_g$, $\gamma\propto N(\Delta \chi)^{3/2}$ and $\kappa \propto N(\Delta \chi)^{3/2}$. As the ratio $\bar{\kappa}/\kappa =0.5$, the Gaussian bending rigidity follows the same law. In an attempt to find the prefactor (which implements finite chain length effects), we note that the mentioned laws are limiting law that apply for long chains. Inspired by finite chain length corrections known for the interfaces in microphase segregation \cite{mocan2018microphase}, we consider prefactors of the type $A + B/\sqrt(N)$. To this end we define $\gamma'=\gamma/(N(\Delta \chi)^{3/2})$ and similarly $\kappa'=\kappa/(N (\Delta \chi)^{3/2})$ and plot these quantities as a function of $1/\sqrt(N)$ in fig \ref{sf3}.  Red dots represent the numerical values estimated from SF-SCF calculations. Blue dashed line is a linear fit. As anticipated the linear fit is acceptable and higher order correction of the type $1/N$ are not needed.   
small maximum and the height of this maximum is a weakly linear function of the chain length $N$. 

Trends observed in Fig. \ref{sf3}a, \ref{sf3}b prove that for the interfacial tension the prefactor grows with decreasing $N$, that is $\gamma \simeq (0.20+0.20\frac{1}{\sqrt{N}}) N\Delta\chi^{3/2}$, whereas the negative slope for mean modulus implies that the prefactor decreases with decreasing $N$: $\kappa \simeq (11.26-10.22\frac{1}{\sqrt{N}}) N\Delta\chi^{3/2}$. At strong segregation for both $\gamma$ as well as for the bending modulus the proportionality constants are independent of $N$. These results reflect on values of $\lambda = \sqrt{\kappa/\gamma}$. Both limits of $\lambda$ is expected feature finite chain length effects that decay with $1/\sqrt{N}$ (shown for weak segregation limit in Fig. \href{4}{4}). These corrections obviously become relatively large for $N$ of order unity. We argue that it is reasonable to find that $\lambda$ is of order segment size $b$: (i) in the strong segregation, where both polymers are in the melt state, the $N$-effects are small. (ii) in weak segregation when the interface is much larger than the coil sizes, the properties of the system should be universal, i.e. not dependent on $N$. 
\begin{figure}[h]
\centering
\includegraphics[width=0.5\textwidth]{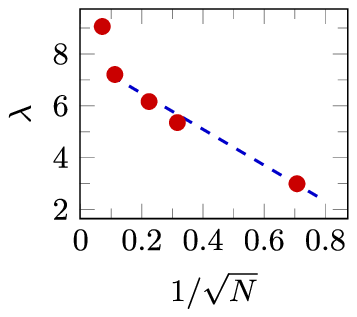}
\caption{Cross-over length $\lambda$ as a function of $1/\sqrt{N}$ at $\Delta \chi = 10^{-3}$ (weak segregation limit).  Fit line (blue dashed) shows that $\lambda_{\rm ws}=(7.86-6.92/\sqrt{N})$ for $N>2$. Fit excludes the $\lambda$ for $N=200$ at $\Delta \chi = 10^{-3}$ did not reach the limiting value $\lambda_{ws}$ (cf. Fig. 2a in the main text).}
\label{sf3a}
\end{figure}

\subsection{The interfacial width and the density difference.}

\begin{figure}[htpb!]
\centering
\includegraphics[width=\textwidth]{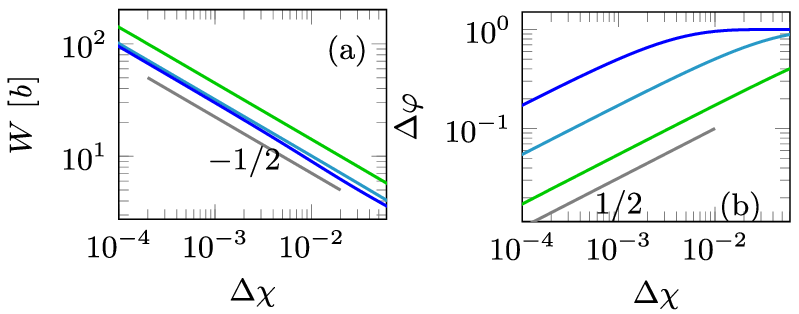}
\caption{(a) Interfacial width  and (b) Density difference as a function of closeness to bulk critical point for various $N$.  Relevant slopes are indicated.}
\label{sf4}
\end{figure}

In Fig. \ref{sf4}a and \ref{sf4}b, the interfacial width and the density difference is presented as a function of $\Delta\chi$ in double logarithmic coordinates for three values of the molecular weights $N=2$, 20 and 200. For the weak segregation regime, that is when the width $W$ exceeds the coil size $R_g$ we find van der Waals like scaling: $W\propto (\Delta \chi)^{-1/2}$ and the density difference obeys to $\Delta \varphi \propto \sqrt{N }(\Delta \chi)^{1/2}$ (Figure not shown). The $\sqrt{N}$ in the latter dependence can be understood from the notion that for weak segregation the van der Waals theory suggests $\gamma \propto (\Delta \varphi)^2/W \propto N (\Delta \chi)^{3/2}$.\cite{safran1994statistical}

\begin{figure}[htpb!]
    \centering
    \includegraphics[width=\textwidth]{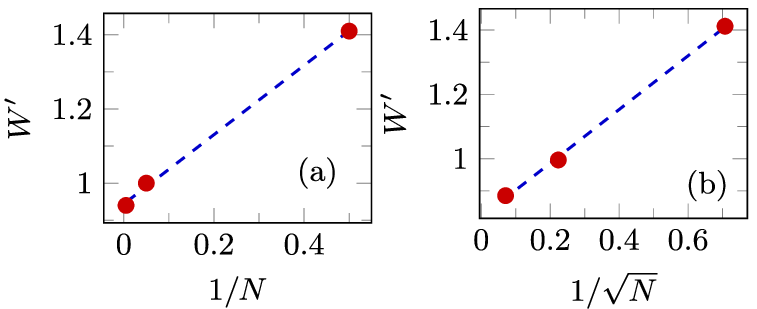}
\caption{(a) $W'$ as a function of $1/N$ in weak segregation ($\Delta\chi < 10^{-3}$). Dashed blue line is a linear fit $W' \simeq (0.94 + 0.93/N)$. (b) $W'$ as a function of $1/\sqrt{N}$ in strong segregation ($\Delta\chi > 5\times10^{-2}$) regime. Blue line is a linear fit $W' \simeq (0.82 + 0.84/\sqrt{N})$. }
\label{sf42}
\end{figure}

Our interest  goes again to the proportionality constants for these scaling dependencies. Interestingly, for the density difference we find for all chain length $\Delta \phi = 1$ in strong segregation and in weak segregation the proportionality constant is independent of $N$: $\Delta \varphi= 1.22  \sqrt{N} (\Delta \chi)^{1/2}$. For $W$ we have non-trivial prefactors both in the weak as well as in the strong segregation limits. Similarly as for the interfacial tension and the rigidities, we define  $W'=W/((\Delta \chi)^{-1/2})$. Using this, we find for weak segregation (cf fig. \ref{sf42}a) that the width of the interface obeys $W = (0.94 + 0.93/N) \Delta\chi^{-1/2}$ while at strong segregation (cf Fig. \ref{sf42}b)  $W =  (0.82 + 0.84/\sqrt{N}) \Delta\chi^{-1/2}$. 

Note that only in the strong segregation case we have the $1/\sqrt{N}$-type finite chain length correction for $W$. This dependence is consistent with the $N$-dependence for the sign switch of $\bar{\kappa}$ (analysed for sufficiently long chains). This leads us to believe that the width of the interface is leading the sign switch of $\bar{\kappa}$. In the text we mention that the Gaussian bending rigidity changes sign when the width of the interface is three to four times the segment size. Numerically the correspondence between the sign switch and the finite chain length correction for the width $W$ do not exactly match. We argue that this is because the sign switch actually takes place in the cross-over regions, that is for short chains it happens in the weak segregation while for longer chains (most of the systems) it occurs in the strong segregation regime. 

\subsection{Interfacial tension and bending rigidities in the absence of non-local interactions. }

The purpose of this section is to prove that the SF-SCF formalism gives essentially the same results as the SCF theory used for microphase segregation, provided that in the SF-SCF calculations we introduce the approximation that $\langle \varphi(r) \rangle \to \varphi(r)$. This approximation is typically used in the SCF theory for microphase segregation, whereas in the SF-SCF theory the non-local interactions are typically taken into account. We focus on the  interfacial tension and the binding rigidities for polymeric interfaces as reported by Matsen \cite{matsen1999elastic}. We will refer to the latter theory as M-SCF, and use m-SF-SCF for the modified SF-SCF formalism.

Both m-SF-SCF as M-SCF need to evaluate the chain partition functions. In both approaches the Edwards diffusion equation is evaluated:
\begin{equation}
\frac{\partial G}{\partial N}=\frac{1}{6} \nabla^2 G - uG
\end{equation}
In the m-SF-SCF this equation is mapped on a lattice using a finite difference approach. The polymers are taken to be built up by segments with ranking numbers $s = 1,\ 2, \cdots, N$ and the space is discretized by a lattice. The lattice site size is taken identical to the segment size so that there is a single length scale $b$ in the problem. The Edwards equation on the lattice results in so-called propagators, which formally implies a change of the chain model from a Gaussian chain (Edwards equation) to a freely jointed chain model with a limited number of step directions. As long as the chains are not strongly stretched the difference between the Gaussian chain model and the freely jointed chain can be ignored. 

In the M-SCF approach the discretization is done using a finite elements approach. Formally the walk along the contour is arbitrarily split up in short contour elements, the number of these elements is independent of the real chain length $N$; the idea is that the chain is seen as an thin featureless thread. In doing so, the results are valid in the infinite chain length limit and the results are typically presented with $\chi N$ as a single parameter. We therefore should expect that m-SF-SCF and M-SCF can only give corresponding results when in m-SF-SCF the chain length is sufficiently large. Without mentioning otherwise we have implemented $N = 200$. Note that in m-SF-SCF the theory will fail to model L/L interfaces in the limit of $N\to 1$.

Both approaches need an iterative method to find the SCF solutions. The 'kitchen' of how this is precisely done is not the topic of this comparison. Assuming that this iterative method is done with the appropriate accuracy, the different approaches should not be relevant. Indeed the good comparison shown below implies that both approaches did what they claimed to do, namely that the accuracy of the SCF solution is sufficient to find numerically accurate results for the mean and Gaussian bending rigidities as well as for the interfacial tension.

\begin{figure}[t]
\centering
\includegraphics[width=\textwidth]{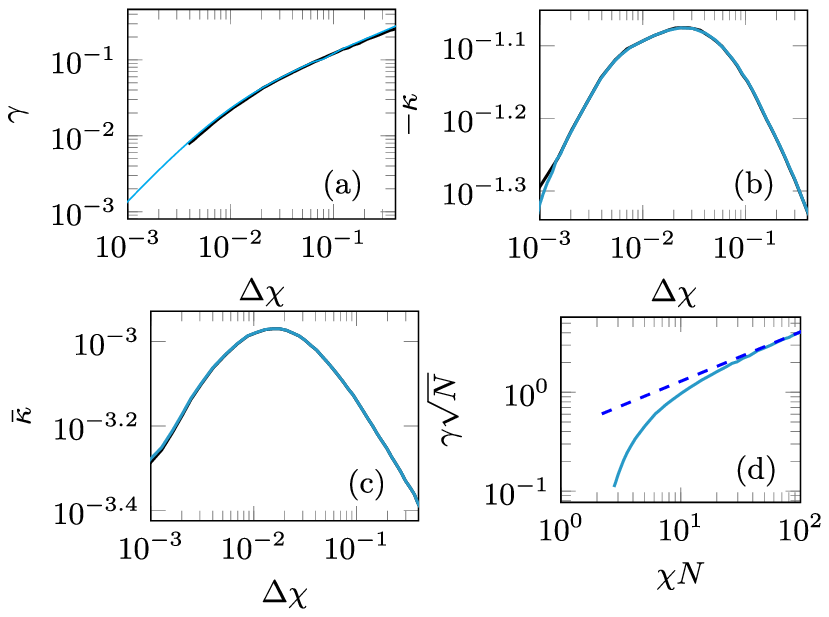}
\caption{Comparison of SCF results for polymeric interfaces when in the SCF theory non-local interactions are not taken into account: (Black line) M-SCF ($N=200$), (Cyan line) m-SF-SCF. (a) The interfacial tension in $k_BT/b^2$. (b) Mean bending rigidity in units of $k_BT$ (c) Gaussian bending rigidity in units $k_BT$ as a function of $\Delta\chi$. (d) Comparison of interfacial tension with analytical strong segregation theory: $\gamma\sqrt{N}$ as a function of $\chi N$. All quantities are plotted in double logarithmic coordinates. All graphs are inspired by similar graphs found in ref \cite{matsen1999elastic} from where also the M-SCF results were extracted.}
\label{sf5}
\end{figure}

In Fig. \ref{sf5} we present the comparison of results for the interfacial tension and the bending rigidities for the m-SF-SCF and the M-SCF theories. When we discuss these results we will contrast these with the full SF-SCF results that are presented in the paper.  

In Fig. \ref{sf5}a we show results for the interfacial tension. As can be seen both approaches match accurately. The power-law scaling at large $\Delta \chi$ reveal the 1/2 slope which is also found in the complete Sf-SCF theory. For weak segregation it is found that $\gamma$ decreases steeper than the 1/2 scaling. Unlike in full SF-SCF where the slope of 3/2 is found, here the results are more leaning to the slope of unity. The fact that near the critical point the mean field results are not expected to be accurate, explains why not much attention was given to the weak segregation results in the microphase segregation community. In Fig. \ref{sf5}d it is shown that for strong segregation the numerical m-SF-SCF as well as the M-SCF approach the strong segregation limiting law predicted by analytical theory (dashed line). 

In Fig. \ref{sf5}b we present results for the mean bending rigidity as a function of $\Delta \chi$ in double logarithmic coordinates. Note the negative sign along the y-axis. Indeed, both in m-SF-SCF as well as in M-SCF the mean bending modulus is \textbf{negative} for all values of $\Delta \chi$. Also in both approaches $-\kappa$ is non-monotonic: the absolute value first increases and then decreases with increasing $\Delta \chi$. These results are in stark contrast with full SF-SCF results for which the mean bending modulus is positive and grows monotonically with $\Delta \chi$, showing two regions of scaling with slopes 3/2 and 1/2 for weak and strong segregation, respectively. 

The Gaussian bending rigidity is presented in  Fig.  \ref{sf5}c. Again m-SF-SCF and M-SCF are in full agreement: $\bar{\kappa}$ is positive for all values of $\Delta \chi$. Similarly as for the mean bending rigidity a non-monotonic dependence is found. The ratio $\bar{\kappa}/\kappa$ is a function of $\Delta \chi$. Again the results are in strong conflict with the full SF-SCF ones for which a sign switch was predicted. Moreover in full SF-SCF both in weak and strong segregation the ratio $\bar{\kappa}/\kappa$ was shown to assume fixed values of $0.5$ and $-1.5$, respectively. These fixed ratio's do not occur in the M-SCF results.

Conclusions regarding the comparison. (i) We have shown that m-SF-SCF predictions for for the interfacial properties of polymeric interfaces did not significantly deviate from the M-SCF predictions. Both approaches use the same way to pin the interface to a specified location which allows the bending at fixed chemical potentials, both approaches use the same Edward diffusion equation and both approaches solve the equations numerically accurate. We see this as a support that also the full SF-SCF results are numerically accurate. (ii) The difference between full SF-SCF and the m-SF-SCF/M-SCF results is large. Results not only differ quantitatively, even qualitatively they do not match. The sign of the bending modulus differs. The functionality with $\Delta \chi$ is completely different. The same applies for the Gaussian bending rigidity. In the full calculations we see a sign switch, which appears to be absent in the m-SF-SCF/M-SCF results. (iii) In full SF-SCF the length scale $\lambda = \sqrt{\kappa/\gamma}$ with a clear physical interpretation, follows meaning-full trends. In m-SF-SCF as well as in M-SCF the value of $\lambda$ can not even be computed because $\kappa$ is negative, and hence $\lambda$ is meaningless. (iv) As the only difference between full SF-SCF and m-SF-SCF rests in the approximation  $\langle \varphi(r) \rangle \equiv \varphi(r)$, we now argue that  $\langle \varphi(r) \rangle$  can not be approximated by $\varphi(r)$. Non-local contributions are essential to find accurate predictions for the (mean field) results near the bulk critical point (for all quantities) and importantly for the mean and Gaussian bending rigidities at strong segregation. 

\bibliography{supplemental}